\documentstyle[12pt,aasms]{article}
\def\kms{km s$^{-1}$}
\def\solar{\ifmmode_{\mathord\odot}\else$_{\mathord\odot}$\fi}
\def\MM{$\cal{M}$\rm}
\def\MSUN{\MM\solar}

\def\etal{et al. }
\def\Ha{H$\alpha$}

\def\kms{km-s$^{-1}$}
\newbox\grsign \setbox\grsign=\hbox{$>$} \newdimen\grdimen \grdimen=\ht\grsign
\newbox\simlessbox \newbox\simgreatbox
\setbox\simgreatbox=\hbox{\raise.5ex\hbox{$>$}\llap
     {\lower.5ex\hbox{$\sim$}}}\ht1=\grdimen\dp1=0pt
\setbox\simlessbox=\hbox{\raise.5ex\hbox{$<$}\llap
     {\lower.5ex\hbox{$\sim$}}}\ht2=\grdimen\dp2=0pt

\begin{document}
\centerline{}

\title{Keck Spectra of Brown Dwarf Candidates and a Precise
Determination of the Lithium Depletion Boundary in the 
Alpha Persei Open Cluster\footnote{Based on observations obtained at the W.M. Keck
Observatory, which is operated jointly by the University of California
and the California Institute of Technology and at the Burrell Schmidt
telescope of the Warner and Swasey Observatory, Case Western Reserve
University} }
\author{John R. Stauffer\footnote{Visiting Astronomer, Kitt Peak National Observatory,
National Optical Astronomy Observatories, operated by the Association of Universities
for Research in Astronomy, Inc., under cooperative agreement with the
National Science Foundation} }
\affil{Smithsonian Astrophysical Observatory, 60 Garden St., Cambridge, MA 02138
     (stauffer@cfa.harvard.edu)}
\author{David Barrado y Navascu\'es} 
\affil{Max-Planck Institut f\"ur Astronomie, Heidelberg, D-69117 Germany
    (barrado@mpia-hd.mpg.de)}
\author{Jerome Bouvier}
\affil{Laboratoire d'Astrophysique, Observatoire de Grenoble,
            Universit\'{e} Joseph Fourier, B.P.  53, 38041 Grenoble Cedex 9,
            France, bouvier@gag.observ-gr.fr (jerome.bouvier@obs.ujf-grenoble.fr)}
\author{Heather L. Morrison\footnote{Heather Morrison is a Cottrell
Scholar of Research Corporation and an NSF Career Fellow}}
\affil{Department of Astronomy, Case Western Reserve University, Cleveland, OH 44106
    (heather@vegemite.astr.cwru.edu)}
\author{Paul Harding}
\affil{Steward Observatory, University of Arizona, Tucson, AZ  85726
   (harding@as.arizona.edu)}
\author{K. L. Luhman}
\affil{Smithsonian Astrophysical Observatory, 60 Garden St., Cambridge, MA 02138
    (kluhman@cfa.harvard.edu)}
\author{Thomas Stanke and Mark McCaughrean}
\affil{Astrophysikalisches Institut Potsdam,
        An der Sternwarte 16,
        D14482 Potsdam
        Germany (tstanke@aip.de; mjm@aip.de)}
\author{Donald M. Terndrup}
\affil{Astronomy Dept., Ohio State University, Columbus, OH 43210
    (terndrup@payne.mps.ohio-state.edu)}
\author{Lori Allen$^2$}
\affil{Smithsonian Astrophysical Observatory, 60 Garden St., Cambridge, MA 02138
    (leallen@cfa.harvard.edu)}
\author{Patrick Assouad}
\affil{International Space University, Parc d'Innovation, Blvd. Gonthier
  d'Andernach, 67400 Illkirch, France (assouad@mss.isunet.edu)}

\vskip1.5truein
\noindent

\begin{abstract}

We have identified twenty-seven candidate very low mass members of the
relatively young Alpha Persei open cluster from a six square degree
CCD imaging survey.  Based on their I magnitudes and the nominal age and
distance to the cluster, these objects should have masses less than 0.1
\MSUN\ if they are cluster members.
We have subsequently obtained intermediate 
resolution spectra of seventeen of these objects using the Keck II 
telescope and LRIS spectrograph.  We have also obtained near-IR
photometry for many of the stars. Our primary goal was to determine 
the location of the ``lithium depletion boundary" and
hence to derive  a precise age for the cluster.   Most of our 
program objects have radial velocities consistent with cluster membership,
moderately strong \Ha\ emission, and spectral
types M5.5 to M8 as expected from their (R-I)$_C$\ colors.  
We detect lithium with equivalent widths greater than or equal to 0.4 \AA\ 
in five of the program objects.

We have constructed a color-magnitude diagram for the faint end of 
the Alpha Persei main sequence, including stars for which high S/N 
spectra in the region of the lithium $\lambda$6708\AA\ absorption 
line have been obtained.  These data allow us to accurately determine 
the Alpha Persei single-star lithium depletion boundary at M(I$_C$) = 11.47, 
M(Bol) = 11.42,
(R-I)$_{C0}$\ = 2.12, spectral type M6.0.  By reference to
theoretical evolutionary models, this converts fairly directly into 
an age for the Alpha Persei cluster of 90 $\pm$ 10 Myr.  That age
is considerably older than most previously quoted ages for the cluster,
but consistent with ages estimated from the upper-main sequence
turnoff using recent models which include a moderate
amount of convective core overshoot.  At this
age, the two faintest of our spectroscopically confirmed members
should be sub-stellar (i.e., brown dwarfs) according to theoretical
models. 

\end{abstract}

\keywords{stars: low mass, brown dwarfs; 
open clusters and associations: individual (Alpha Persei) }

\section{Introduction}

Open clusters provide the best means to calibrate stellar age scales.  
Traditionally, this has meant comparing a color-magnitude diagram of 
the upper main sequence and red giant branch of a given cluster to 
theoretical evolutionary isochrones for high mass stars (cf. Sandage 
1957; Patenaude 1978; Meynet \etal 1990).  While this method is
certainly valid qualitatively, it is subject to uncertainties with
respect to the rotation rates and duplicity of the upper-main sequence
turnoff stars, problems related to the fact that the form of the
initial mass function insures that in most cases the number of stars
defining the turnoff region will be small, and - perhaps most
importantly - the uncertainty regarding the amount of mixing of
hydrogen rich matter into the convective core (normally parameterized
as ``convective core overshoot").  Uncertainty in this latter issue 
has led to differences in the quoted ages of such well-known clusters
as the Pleiades and Hyades by up to a factor of two, with the age
of the Pleiades in particular ranging from $\sim$75 Myr to $\sim$150
Myr (Mermilliod 1981; Mazzei \& Pigatto 1988).

An alternative means to estimate open cluster ages is to use the
location of the pre-main sequence isochrone instead.  This has the
advantage of completely avoiding the uncertainty in the convective
core overshoot parameter and also takes advantage of the form of
the IMF (i.e., there are expected to be many more low mass, pre-main
sequence stars than high mass post-main sequence stars in a
typical open cluster of age $\leq$ 200 Myr).  Of necessity, this method
can only be applied to young, nearby clusters, where the pre-main sequence
turn-on point occurs at relatively bright absolute magnitudes and where
the apparent magnitude of the stars defining the pre-main sequence
isochrone are within the range accessible for observation.   For
ages older than about 50 Myr, the method becomes difficult to apply because
the displacement of the pre-main sequence isochrone above the ZAMS
becomes small when compared to the possible systematic errors in 
locating the cluster isochrone relative to an appropriate ZAMS relation.
Recent examples of the application of this method to derive age
estimates for NGC2547 and IC~2391 are provided in Jeffries \& Tolley
(1998) and Stauffer \etal (1997a), respectively.

A new way to estimate the age of open clusters has recently been
proposed by Basri, Marcy \& Graham (1996=BMG), Bildsten \etal (1997),
Ushomirsky \etal (1998),
Ventura \etal (1998) and others.  The idea behind this method is that
for stars near the substellar mass limit, the age at which stars become
hot enough in their cores to burn lithium is a sensitive function of
mass.  Furthermore, it is argued that the physics required to predict
the location of the ``lithium depletion boundary" as a function of age
is very well understood and not subject to significant uncertainty
(cf. Bildsten \etal 1997).  In the mass range of interest, stars
are fully convective, and the core lithium abundance will be directly
reflected in the surface lithium abundance - and the latter can be
be determined by use of the 6708\AA\ \ion{Li}{1} doublet.  Because the
atmospheres of these quite cool stars are complicated (Allard
\& Hauschildt 1995), it is not yet possible to derive accurate
lithium abundances from such spectra even if the lithium doublet
can be detected at high signal-to-noise.  However, the lithium
depletion boundary is expected to be very sharp, with essentially no
lithium just above the boundary and essentially cosmic lithium
abundance just below the boundary.  Therefore, it should suffice to 
simply detect the point at which the lithium doublet is solidly 
detected in order to derive a precise age for the cluster.

BMG and Rebolo \etal (1996) were the first to successfully apply
this test by detecting lithium in three substellar objects in the
Pleiades.  However, it was later determined (Basri \& Mart\'{\i}n 1998)
that the exact location of the lithium boundary (and hence the
Pleiades age)  was uncertain because the brightest of the three
objects (PPL15) is a nearly equal mass binary.  This problem was 
resolved by Stauffer, Schultz \& Kirkpatrick (1998=SSK), who obtained
spectra of an additional 10 faint Pleiades members, five of which
still retain their lithium.   By providing measurements of a large 
number of stars near the lithium boundary, SSK were able to determine
the absolute magnitude of the lithium depletion boundary to an accuracy
of about 0.1 mag, corresponding to an age uncertainty of about 8 Myr.
The age derived for the Pleiades by SSK was 125 Myr.

Basri \& Mart\'{\i}n (1999=BM99) recently reported Keck HIRES 
spectra of the faintest several candidate Alpha Per members from Prosser 
(1994).  They detected lithium in one of those stars (AP270, I$_C$ $\simeq$
17.9), failed to detect it in another somewhat brighter star thought to be
a cluster member (AP275, I$_C$ $\simeq$ 17.2), and also failed to detect
lithium in a star of uncertain membership status with luminosity intermediate
between the other two stars (AP281, I$_C$ $\simeq$ 17.6).  Based on these data, 
BM99 determined the cluster age to be in the range 58 to 90 Myr. 
We have now obtained spectra of a large number of new
candidate Alpha Per low mass members;
we report the results of the analysis of those spectra
here and provide a new, more precise age estimate for the cluster. 
In a separate paper, we report the probable identification of
the lithium depletion boundary in an even younger cluster, IC~2391
(Barrado y Navascu\'es \etal 1999).

\section{Observations and Data Reduction}

\subsection{Identification of Candidate Faint Members from a CCD Imaging Survey} 

The primary published source lists for low mass members of Alpha Per are
papers by Stauffer \etal (1985, 1989) and Prosser (1992, 1994).  Most of the
proposed members in these papers were identified from proper motion data;
however, the faintest candidates were selected solely on the basis of
having locations in a V vs. V-I color-magnitude diagram that are consistent
with cluster membership, with followup low-resolution spectroscopy when
possible.  
BM99 observed the faintest published candidate members of Alpha Per.
In order to determine a more precise age for the Alpha Persei cluster and 
to identify truly substellar cluster members, it was necessary for us to create
a new list of fainter candidate cluster members.  We have addressed that
need by obtaining imaging photometry in R and I of about six square degrees of the
cluster deep enough to allow us to identify cluster members to 
I$_C$ $\simeq$ 19.   The telescopes
we used and the individual areal coverage are indicated in Table 1.  The
last dataset indicated in Table 1 was obtained too late to be included
in the preparation for the Keck spectroscopy run, and we have not 
included those data in estimating the areal coverage attained for the
current program.

The CCD frames obtained by us fell within a region approximately bounded by
03h15m $<$ RA(2000) $<$ 03h40m  and 48${\deg}$\ $<$ DEC(2000) $<$ 
50${\deg}$30'.  
For each of the datasets
described in Table 1, we used standard tools within IRAF
\footnote{IRAF is distributed  by National
Optical Astronomy Observatories, which is operated by
the Association of Universities for Research
in Astronomy, Inc., under contract to the National
Science Foundation, USA}
to calibrate
the images, and we used DAOPHOT aperture photometry to obtain magnitudes
for the stars in each region.  Standard stars from Landolt (1992) were
observed to place the photometry on the Cousin's photometric system.
In most cases, the internal errors in our photometry for the stars
of interest should be of order 0.05 mag (1$\sigma$) or less.  The external
errors may be larger due to difficulties in placing our photometry on
the Cousin's system (our program stars are redder than any of the Landolt
standard stars used, in most cases) and due to the expected variability
of the program stars (due to starspots).   Comparison of multiple observations
for a few stars suggests an external accuracy of order 0.1 mag for
our photometry.

Candidate Alpha Persei cluster members were defined as those falling
above a ZAMS line shifted to the distance and reddening of Alpha Per
(assumed to be (m-M)$_o$ = 6.23, A$_V$ = 0.30 - cf. Pinsonneault
\etal 1998; A$_{I}$ = 0.17; E(R-I)$_C$\ = 0.07).  
In practice, this corresponds to simply selecting the
reddest stars in each field in the I magnitude range of interest.
Because our primary goal was simply to identify a set of good candidate 
cluster members near the lithium depletion boundary, we did not 
attempt to insure that our search was complete nor did we push our
object identifications to the faintest limit possible.   In a separate
paper, we will report the results of a more complete, deeper search
for Alpha Persei members (Bouvier \etal 1999).

We illustrate the technique used to select candidate Alpha Persei
members in Figure 1, where we show the I vs. (R-I)$_C$\ color-magnitude
diagram for one of the fields observed with the CWRU Burrell Schmidt telescope.
The ZAMS shown in Figure 1 is the same as that utilized
in Bouvier \etal (1998).  We have only retained stars for
further consideration with 16.25 $<$ I$_C$\ $<$ 18.75 and with R-I
colors such that they are displaced significantly above (or redward)
from the ZAMS due to our
narrow focus on defining the lithium depletion boundary
in Alpha Per - other Alpha Per members are presumably
present in Figure 1 both brighter and fainter than
this range, but were not of interest here.  The same
method was applied to each of our datasets, yielding a
total of 27 candidate Alpha Per members.  A color-magnitude
diagram for the entire
set of candidate faint Alpha Per members is shown in
Figure 2, where we also include the faintest candidate Alpha Per
members from Prosser (1992, 1994).  Table 2 provides 
coordinates and magnitudes for these stars; finding
charts are provided in the appendix.  The coordinates were derived
either using the ``ccmap" routine in IRAF (with previously known Alpha
Per members providing the input data) or by using the world coordinate
system imposed on the image by the telescope operating system with a
zero point correction determined from cataloged Alpha Per members.
Based on the rms provided by the ccmap routine and checks we have
made when we have more than one position estimate for a given star,
we expect the positions to be accurate to better than 2 arcseconds.
The names for our candidates are a continuation of the series
used by Stauffer \etal (1984,1989) and Prosser (1992, 1994), except
that we have left a short gap between the last catalog entry of
Prosser (1994) and our first candidate.  We have left the gap in
case one or two candidate Alpha Per members have been proposed by
others between 1994 and 1999, though we are unaware of any such
candidates.  Because the AP catalog includes stars which were later
found not to be Alpha Per members, we do not believe that the
existence of this gap in the numbering system should have any
deleterious effect.

\vskip0.1truein

\subsection{Spectroscopic Observations}

We obtained spectra for some of our Alpha Per candidate
members on September 6 and November 24-26, 1998, using
the Keck II telescope and the
Low-Resolution Imaging Spectrograph (LRIS) (Oke \etal
1996).  We used the 400 l/mm grating to obtain quick,
low-resolution spectra in the wavelength range
$\lambda\lambda$6250-9900\AA\ in order to identify
probable field star interlopers in our sample based
on their lack of \Ha\ emission or spectral type
inappropriate for our measured photometry.  Integration
times for these spectra were typically of order eight
minutes.  For the most promising candidates, we 
subsequently obtained higher
signal-to-noise spectra using the 1200 l/mm grating,
with typical integration times of order 90-120 minutes.
The wavelength range covered with these spectra was
$\lambda\lambda$6430-7650\AA.   The spectral resolutions
for the two gratings were 7.0\AA\ and 2.5\AA, respectively.

In total, we obtained low-resolution spectra of 14 of our
candidate Alpha Per members, and high resolution spectra
of 11 of these stars.
We also obtained spectra of a set of M dwarf spectral
standards and a small number of brighter candidate
Alpha Per members selected from Prosser (1992,1994).

In a previous paper (SSK)
we have analysed Keck spectra of a set of Pleiades brown
dwarf candidates in order to define the lithium depletion
boundary in that cluster.  We have applied the
same analysis techniques to our new Alpha Per spectra.
\Ha\ equivalent widths
were determined using the SPLOT utility in IRAF via both
Gaussian fits and direct integration over the profile.
Radial velocity estimates from the \Ha\ fits
were determined by using nearby
OH emission lines in the sky spectrum to provide an in-situ
flexure correction to the wavelength scale.    Based on
our previous experience using this technique (cf. SSK) and 
on the scatter of the derived radial velocities for the
Gliese M dwarfs we observed, we believe these radial
velocities should have 1$\sigma$\ errors of 5-10 \kms.
A number of
indices sensitive to the continuum slope and to molecular
band strengths (TiO, VO, CaH) were calculated for each spectrum
in order to yield accurate estimates of the intrinsic
(R-I)$_C$ color for the program objects, using the Gliese
catalog M dwarfs to calibrate the indices.  The primary
indices measured for this purpose were PC2, a narrow-band
color between 7000\AA\ and 7500\AA\ (Mart\'{\i}n \etal 1996);
the strength of the VO band at 7440 \AA\ (Kirkpatrick
\etal 1995); and a measure of the height of the continuum
peak at 8100\AA\ relative to the molecular ``valleys" on
either side which we call `C81'.  We have defined this
index as:  C81 = 2.0 * f81/(f79+f85), where f79, f81 and f85 are 
the summed counts in a 50\AA\ bandpass centered on
7890\AA,  8140\AA\ and 8515\AA, respectively.
We also measured equivalent widths for the NaI
doublet at 8200 \AA\ for stars observed with the low
resolution grating, and the \ion{Li}{1} 6708 \AA\ doublet for
stars measured with the high resolution grating.  Table 3
provides the results of our spectroscopic data analysis.

Figure 3 provides a montage of the high resolution spectra
of four of our target stars where we have detected lithium,
in order to illustrate that the $\lambda$6708\AA\ lithium detection 
is reliable.  Figure 4 shows the low resolution
spectra of our two latest type candidate Alpha Persei
members, which according to the discussion in the next
section should have masses below the substellar boundary
for the inferred age of Alpha Persei.   We estimate the spectral types
of these stars to be M7.5 and M8.

\subsection{Infrared Photometry}

We have obtained K-band photometry for most of the objects
observed spectroscopically using near-IR cameras attached to
the Keck-I, Calar Alto 3.5m, IRTF and Mt. Hopkins 1.2m
telescopes.  In all cases, magnitudes were derived from
aperture photometry.  Standard stars were selected either from
the faint-standards list of Persson \etal (1998) or the UKIRT
faint standards list of Casali (1992).
The nominal 1$\sigma$ photometric errors were usually of
order 0.05 magnitudes or slightly better.   One object -
AP306 -- had two very discrepant measurements obtained
at widely different times; we have adopted the fainter of
the two measurements on the assumption that the other 
observation may have been obtained while the star was
flaring.

The K band photometry is provided in Table 2.

\section{Discussion}

\subsection{Cluster Membership}

Before we can consider the implications of the new data, we
must first establish which of the stars we have observed are in
fact probable members of the Alpha Persei cluster.  Traditionally,
cluster membership comes primarily from proper motions and
secondarily from radial velocities, color-magnitude diagrams,
color-color diagrams and spectral indices (metallicity, chromospheric
activity, lithium abundance, etc.).  For the magnitude range of
interest to us, proper motions are not possible at present due to
lack of suitably deep first epoch images.  In any case, proper motions
are only of limited use in Alpha Per because the cluster motion
is not significantly displaced from the centroid of field star
motions in that direction (cf. Prosser 1992).  By selection,
all of our stars are consistent with being cluster members based
on their location in a color-magnitude diagram using the photometry
we derived from our CCD images.   How do the other data we have
obtained affect the assessment of cluster membership for these stars?

Figure 5 shows two color-magnitude diagrams for our program
stars.   For the I$_C$ vs. (R-I)$_C$ diagram, we have used the
spectroscopically estimated (R-I)$_C$ color rather than the
color from aperture photometry because we believe the former estimate
provides a more accurate and homogeneous measure of the star's color,
and one that is reddening free.   Two stars - AP312 and AP314 - fall 
well below the lower envelope of the locus of most of the program 
objects in both diagrams, and we consider them to be nearly certain non-members.
AP314 is also the only object observed which does not have \Ha\ in
emission, thus adding confidence that it is in fact a much older,
field dM star.  One other star - AP322 - falls a few tenths of a 
magnitude low in both figures, and has the most discrepant radial
velocity of all of the stars observed at high resolution.   We 
consider it to be a probable non-member.  AP317 falls well below
the locus of the other stars in the I-K diagram, but within the
main locus in the R-I diagram - either the measured K magnitude is 
about 0.4 mag too faint or this is likely to be a background, reddened
field star; we consider it to be just a possible member.
Finally, two stars - AP301 and AP325 - are displaced well above the 
locus of the others in both diagrams; because their spectral properties
are consistent with cluster membership, we consider them to be
probable binary members of Alpha Per.

With the exception of AP314, all of the program objects have \Ha\ in
emission with equivalent widths ranging from 3 to 10 \AA.  We expect
the \Ha\ emission equivalent widths in the late M dwarf members of Alpha 
Per to be like that found for similar stars in the Pleiades
and Hyades based on evidence that the rotational velocities for these
stars should be quite high (Jones, Fischer \& Stauffer 1996;
Stauffer \etal 1997b; Oppenheimer \etal 1997).    Many field dM
stars in this spectral type range also have \Ha\ emission with these
strengths, so this is a necessary but not sufficient condition for
cluster membership.  Field dM stars in this spectral type range
with the lithium 6708\AA\ doublet strongly in absorption are, however,
extremely rare.  We therefore argue that the five program objects
with prominent lithium absorption features are nearly certain Alpha
Per members.  The radial velocities
we have derived are consistent with Alpha Per membership for all of
our targets except for AP322 given that $<$v$_r$$>$ $\simeq$ -2 \kms\
for Alpha Per (Prosser 1992) and that our 1$\sigma$\ measurement
error is 5-10 \kms. 
Finally, the fact that the NaI 8200\AA\ doublet equivalent 
widths for our program stars are greater than 5 \AA\ indicates that 
they are neither giants nor $\sim$1 Myr old PMS stars.  When
compared in detail with the small number of Gliese M dwarfs we
have observed, the NaI equivalent widths for the Alpha Per members
are slightly smaller than for Gliese M dwarfs of the same spectral
type.  A similar effect has been noted in the Pleiades by Steele
\& Jameson (1995) and Mart\'{\i}n \etal (1996) and in IC~2391 by
Barrado y Navascu\'es \etal (1999); those authors attribute the
result to lower gravity for the slightly PMS cluster stars.  This
is plausibly the right explanation for our Alpha Per sample also,
and thus argues that we have indeed isolated a set of young M dwarfs.

Given the above considerations, we believe that all but a few of the
stars we have observed spectroscopically are probable Alpha Per
members.  The likely non-members are AP312, AP314 and AP322, and we
will exclude those stars from further discussion.  AP317 has
uncertain membership, and in the remaining discussion we will consider it
as only a possible member.

\subsection{The Lithium Depletion Boundary and the Age of Alpha Per}

We are now in a position to obtain our estimate of the
location of the lithium depletion boundary.  Figure 6 shows an
I$_C$ vs. (R-I)$_C$ plot of our probable cluster members, where
we highlight the stars where lithium has been detected. 
We also include AP270 in the plot, since AP270 was
the one star with a lithium detection from Basri \& Mart\'{\i}n (1999).
As before, the x-axis is derived from our spectra and so is reddening
free; the y-axis has been corrected for a uniform assumed
extinction of A$_I$ = 0.17.  The figure shows a clearly delineated boundary between
the stars with and without lithium.   The Alpha Per lithium boundary
occurs at (R-I)$_C$ = 2.12, with an uncertainty of only a few
hundredths of a magnitude, largely dependent on possible systematic
errors in the conversion from spectral indices to intrinsic R-I color.
One concern might be that the gravity difference between the slightly PMS
Alpha Per stars and the main-sequence Gliese M dwarf calibrators might lead
to a different relation between spectral index and color, thus causing our
inferred R-I colors to be systematically in error.  We believe that
the actual errors from this cause are small for several reasons:
(a) our spectra
cover essentially exactly the wavelength region sampled by the R and
I photometric bands, and so whatever is acting to affect the broadband
colors will also be affecting our spectra; (b) our spectral indices
include both measures of local continuum color - e.g. the PC indices
from Mart\'{\i}n \etal (1996)- and measures of the depth of molecular
bands.  There is good agreement between the colors inferred from both
types of indices; (c) the expected gravity difference between the 
Alpha Per stars and ZAMS star at the 
color of the boundary is not large - using the 
Baraffe \etal (1998 = BCAH) models and the age we derive below, log g for
a star at the boundary at Alpha Per age is
$\sim$4.85 vs. 5.2 on the ZAMS; 
and - perhaps most importantly - (d) we have a test of our
assumption from the Bouvier et al. (1998) and Stauffer et al. (1998) Pleiades
papers.  That is, Bouvier et al. 
obtained photometric (R-I)$_C$\ colors for a set of very low mass Pleiades stars, 
and Stauffer et al. measured spectroscopic estimates of
(R-I)$_C $\ in exactly the same way we have for Alpha Per.  After correcting
for the small and nearly uniform reddening in the Pleiades, the mean
difference between the photometric and spectroscopic estimates of
R-I is only 0.01 mag (for seven stars).  Since the Pleiades is only 
slightly older than Alpha Persei, we believe this demonstrates that
the spectroscopic R-I estimates we derive in Alpha Per are not significantly
in error due to the slight gravity difference compared to the calibrators.
There is similarly good agreement between the photometric and spectroscopic
measures of (R-I)$_C$\ in IC~2391 (Barrado y Navascu\'es \etal 1999).

The R-I color of the lithium boundary that
we have derived for Alpha Per is $\sim$0.08 magnitudes bluer than for the
Pleiades (see SSK), as qualitatively expected given the assumed younger age for
Alpha Per.  In order to estimate the I magnitude of the lithium
depletion boundary, we have
fit an illustrative single-star locus to the lower envelope of the
observed stars, as indicated by the dashed curve in Figure 6.
Using this curve as a guide, we estimate that the lithium depletion
boundary is at I$_{C,0}$ $\sim$ 17.70.  The boundary is very well
defined in terms of the R-I color, but more poorly defined in terms
of the I magnitude - presumably due to a combination of differential
reddening, unresolved binaries and measurement errors.  We adopt
0.15 mag as a somewhat arbitrary estimate of the possible one sigma
error in our location of the I magnitude of the lithium depletion
boundary.  

To derive an age, we must adopt a distance to the Alpha Per cluster.
Based on his analysis of the existing membership studies and photometry
available for Alpha Per, Prosser(1992,1994) advocated a distance modulus
of (m-M)$_o$ = 6.15.   The most recent analysis of the Hipparcos data
for the Alpha Per cluster yields a distance modulus of (m-M)$_o$ = 6.31
(van Leeuwen 1999).
Pinsonneault \etal (1998) derived (m-M)$_o$ = 6.23 based on intercomparing
main-sequence fits for the open clusters analysed by the Hipparcos teams.  
We adopt
(m-M)$_o$ = 6.23 as a convenient average of the above results,
with the realization that there is an uncertainty of order 0.1 mag in
this estimate.

With the above distance, the absolute I magnitude of the lithium
depletion boundary in Alpha Per is M(I$_C$) = 11.47.   Figure 3 of our paper
on the lithium depletion boundary in the Pleiades (Stauffer \etal
1998) - based on the models of Baraffe \etal (1998) - provides a direct
calibration to convert the absolute I magnitude of the lithium boundary to an
age for a cluster.  Using that calibration, we derive an age for
the Alpha Persei cluster of 90 Myr.  Adding the uncertainty in the Alpha 
Per distance and
the uncertainty in the measured location of the lithium depletion
boundary in quadrature, we crudely put an error bar of 0.18 mag on our
estimate of the absolute I magnitude of the lithium depletion boundary.
Folding that error into Figure 3 of Stauffer \etal (1998), 
the uncertainty in our age estimate from this method is approximately $\pm$8 Myr.

Because the lithium depletion boundary is much more precisely defined
in terms of its R-I color, it would be useful to be able to use the
BCAH model predictions of the R-I colors as a function of age to derive
an age estimate for Alpha Persei.  Unfortunately, the current status
of the models precludes that - for the cool dwarfs of interest, the
models do not yet accurately reproduce V-I or R-I colors, presumably due
to inaccuracies in the molecular line lists.  It is believed
that the BCAH models are accurate at K, and so we can derive alternate
age estimates using both the absolute K magnitude of the lithium boundary
and the I-K color of the boundary.  We are limited in our ability to
do that because there are fewer stars with
measured K photometry and because the I and K photometry are
not contemporaneous, which is significant since we expect most of these stars
to be photometric variables.
As can be inferred from Figure 5b and Table 2,
the lithium boundary in the K vs. I-K diagram is defined by
AP323 (the bluest star in the diagram with lithium, but probably
a binary), AP310 (the
reddest apparently single, Alpha Per member without lithium in
this diagram), and AP300 (the bluest apparently single member with
lithium).  Our best estimate is that the lithium boundary is at
about M(K) $\sim$\ 8.3 and (I-K)$_0$ $\sim$\ 3.07.  Those values correspond
to ages of about 80 Myr and 100 Myr, respectively, using the
BCAH models.

Finally, it is also possible to use the theoretical models by
D'Antona \& Mazzitelli (1997) or Burrows \etal (1997) to estimate
the cluster age from the location of the lithium boundary.  For these
models, we convert the I magnitude of the boundary to a bolometric
magnitude using an average of the bolometric corrections inferred from
Monet \etal (1992) and Leggett \etal (1996).   We derive
M(Bol) $\sim$ 11.40 in that fashion. Using that bolometric magnitude
and the theoretical models we can then estimate the age.
The resultant ages for the two models are $\simeq$ 85 Myr for
D'Antona \& Mazzitelli and about 90 Myr for the Burrows \etal models.

From the above considerations, we believe the best current estimate
for the age of the Alpha Persei cluster based on our lithium data
and the current models is 90 $\pm$ 10 Myr.  An improved estimate
will be possible when either additional photometry becomes available
(in particular, more and better near-IR colors) or when the theoretical
models can be improved to predict better the R-I or V-I colors
of stars in this spectral type range.

As noted previously, BM99 derived an age estimate for Alpha Persei
using the lithium depletion boundary method of 65 Myr.   However, this
was essentially the minimum age for the cluster that was compatible
with their data. They quoted a maximum age
compatible with their data of $\sim$90 Myr, corresponding to locating
the lithium depletion boundary essentially at the magnitude of the
one star where they detected lithium (AP270).  We believe,
based on our larger sample of stars, that the Alpha Per lithium
depletion boundary is in fact essentially just at (or slightly
brighter than) AP270, and therefore there is little real disagreement
between the two results.

\subsection{Substellar Members of the Alpha Persei Cluster}

If we adopt an age of 90 Myr for Alpha Persei, then the mass corresponding
to the lithium depletion boundary is $\sim$0.085 \MSUN\ according to the
BCAH models, and an object currently
at that absolute magnitude is destined to be a hydrogen burning, main sequence
star.  At that age, the I magnitude corresponding to 0.075 \MSUN\ according
to their models is $\sim$18.2 (at the Alpha Per distance and reddening).  
Using the D'Antona \& Mazzitelli (1997) models, 
we estimate this point to be at I$_C$\ $\sim$ 18.0.
Taking the fainter of the two estimates (the BCAH estimate), we infer that 
the faintest two
of our probable members for which we have both spectroscopic and photometric
information are likely to be substellar.  The
inferred mass of the faintest of these stars (AP326)
based on the BCAH models is $\sim$ 0.063 \MSUN.

\section{Summary and Implications} 

We have used an imaging survey of six square degrees of the Alpha Persei
cluster to identify about a dozen highly probable cluster members
with estimated masses near the hydrogen burning mass limit.  Spectra
and photometry of these stars has allowed us to determine the
location of the lithium depletion boundary in the cluster to be
at M(I$_C$) = 11.47, corresponding to an age of 90 Myr according to
the theoretical models of Baraffe \etal (1998).  The two
coolest probable members for which we have spectra have masses
below the hydrogen burning mass limit for this age, and are
therefore brown dwarfs.

All three clusters with ages derived from the lithium depletion
boundary (Pleiades - Stauffer \etal 1998; Alpha Persei; and 
IC~2391 - Barrado y Navascu\'es \etal 1999) have lithium ages which
are significantly older than that which would be derived from theoretical
models which do not include convective core overshoot at high masses.
The lithium depletion ages for the Pleiades and Alpha Persei 
are in good agreement, however, with upper-main sequence turnoff 
ages derived from theoretical models by Ventura \etal (1998) which 
incorporate convective core overshoot and modern opacities (they derive ages
of 120 Myr and 80 Myr for the Pleiades and Alpha Persei, respectively).
The ratio of the age derived from the lithium boundary
to that derived from the upper-main sequence turnoff
using older non-overshoot models 
(e.g., Mermilliod 1981)
is approximately the same in all three cases, suggesting a similar 
amount of core overshoot is needed to normalize the ages - and hence
that the amount of convective core overshoot is not a strong function
of mass.

It will be important to extend the determination of the lithium depletion
boundary ages to clusters of other ages in order to continue to refine
the open cluster age scale and to constrain the mass dependence of the
convective core overshoot parameter.  The amount of effort required
to do this should not be underestimated, however.  
For a cluster older than those where the boundary has so far been
determined, the absolute magnitude of the lithium depletion boundary
is fainter, making spectroscopy more difficult.
The Hyades - at an age of 600 Myr or older - is too old for the test 
to be applicable because the lithium boundary becomes fixed at
$\sim$0.06 \MSUN\ for all clusters older than about 250 Myr (i.e.,
the cores of brown dwarfs less massive than that never become hot
enough to burn lithium).  
Given the realities of what clusters exist within 500 pc from the Sun, 
perhaps the best target cluster for the purpose would be NGC2516 with
a nominal distance of order 380 pc and a nominal age of about 140 
Myr (Jeffries, Thurston \& Pye 1997).  Assuming the real age is
slightly older (say 150 or 160 Myr), the lithium depletion boundary
would be at about I $\sim$ 20.6 according to the BCAH  models,
which is faint enough that detection of the lithium $\lambda$6708\AA\
feature would be a challenge even with a 10m class telescope.
At the other end of the age range,
the lithium depletion boundary method is not applicable for very
young clusters (ages less than about 10 Myr) because lithium burning
will not have begun throughout the mass range of interest.  For
clusters between 10 and 50 Myr old, the lithium depletion boundary
should occur at relatively bright absolute magnitudes, but successful
implementation of the method will still not be
a simple exercise.  The primary difficulty is a lack
of known, or at least known and well-studied, clusters in this age range
and the concomittant need to do
a significant amount of ``spade" work.  We nevertheless expect that
the work in this area will bear fruit in the near future, and that
other lithium depletion ages will become available for clusters younger
than IC~2391.

\acknowledgments 
The spectra reported here were obtained at the W. M. Keck Observatory. The
W. M. Keck Observatory is operated as a scientific partnership
between the California Institute of Technology and the University            
of California.  It was made possible by the generous financial
support of the W.M. Keck Foundation.
JRS acknowledges support from NASA Grants NAGW-2698 and NAGW-3690.
DBN acknowledge the fellowship by the {\it Instituto
Astrof\'{\i}sico de Canarias}, Spain, and the 
{\it Deutsche Forschungsgemeinschaft}, Germany.
DMT and HM thank the NSF for support through NSF Grants AST-9731621 
and AST-9624542.
We thank Lee Hartmann, Cesar Brice\~no, Michael Pahre, Andrew Connelly, Ben Oppenheimer,
Andrea Ghez and Russel White for obtaining data used in the preparation
of this paper.  We also thank NASA for making the Keck telescopes
available to the community for this type of research.

\appendix

\section{Finding Charts for New Candidate Alpha Persei Cluster Members}

We provide finding charts for our new Alpha Persei candidate members
in Figure 7.  Each chart is 3' x 3', with North to the top and East to
the left.  The candidate member is indicated by a circle.  The charts
were created from several different telescope/camera systems, and so
the pixel scale varies from chart to chart - but the field of view
is the same in all cases.
\vfill
\eject

\vfill
\eject
\centerline{FIGURE CAPTIONS}
\vskip0.1truein
\noindent
Figure 1:  Color-magnitude diagram for an approximately 1.5
square degree region of the Alpha Persei cluster based on data
from the CWRU Schmidt telescope.  The vast majority of the
stars are background dwarfs in the galactic disk.  The large
solid dots indicate objects that we have identified as possible
cluster members.  The solid line is an empirical main sequence
shifted to the distance (r=176 pc) and reddening of the cluster
(A$_I$ = 0.17; E(R-I)$_C$\ = 0.07).
\vskip0.1truein
\noindent
Figure 2:  Color-magnitude diagram for all of the candidate
Alpha Per members identified in our imaging survey.  Also shown
are a number of probable Alpha Per members identified by
Prosser (1992,1994).
\vskip0.1truein
\noindent
Figure 3:  A small portion of the Keck LRIS 1200 l/mm spectra
for four of the Alpha Per members where lithium was detected.  
The $\lambda$6708\AA\ feature is identified by the arrow.
The Alpha Per member is shown as a solid line; a scaled
spectrum of a field M dwarf of the same spectral type with
no detected lithium is overplotted as a dashed line.
\vskip0.1truein
\noindent
Figure 4:  Low resolution spectra of our two latest type
probable Alpha Per members.  AP326 has spectral type $\sim$M7.5
and AP306 has spectral type $\sim$M8; their
inferred masses are of order 0.065 \MSUN.
\vskip0.1truein
\noindent
Figure 5a:  I$_{C}$ vs. (R-I)$_C$ plot for our Alpha Per sample.
The y-axis is the observed I magnitudes from our photometry,
whereas the x-axis values are the intrinsic (R-I)$_C$\ colors
estimated from our spectra.
\vskip0.1truein
\noindent
Figure 5b: K vs. (I-K) plot for the stars with such data in
our sample.  
\vskip0.1truein
\noindent
Figure 6:  I$_{C0}$ vs (R-I)$_C$ plot for the Alpha Per
stars in our sample believed to be probable or possible
members.  The I magnitudes have been corrected for an
assumed uniform extinction of A$_I$\ = 0.17.  The (R-I)$_C$\
colors are our spectroscopic estimates and so are inherently
reddening free.  Also included are probable members from
Prosser (1992,1994).  The different symbol types are
identified within the body of the figure.  The solid curve
is our empirical main sequence, and the dashed
curve is a plausible single-star Alpha Per locus near the
region of the lithium depletion boundary.  AP270 is plotted
using photometry from Prosser (1994), with the measured V-I
color transformed to R-I using data in Leggett (1992).  
The assumed distance modulus is (m-M)$_o$\ = 6.23.
\vskip0.1truein
\noindent
Figure 7:  Finding charts for the new candidate Alpha Persei
members from Table 1.

\begin{figure*}
\vspace{18cm}

\includegraphics{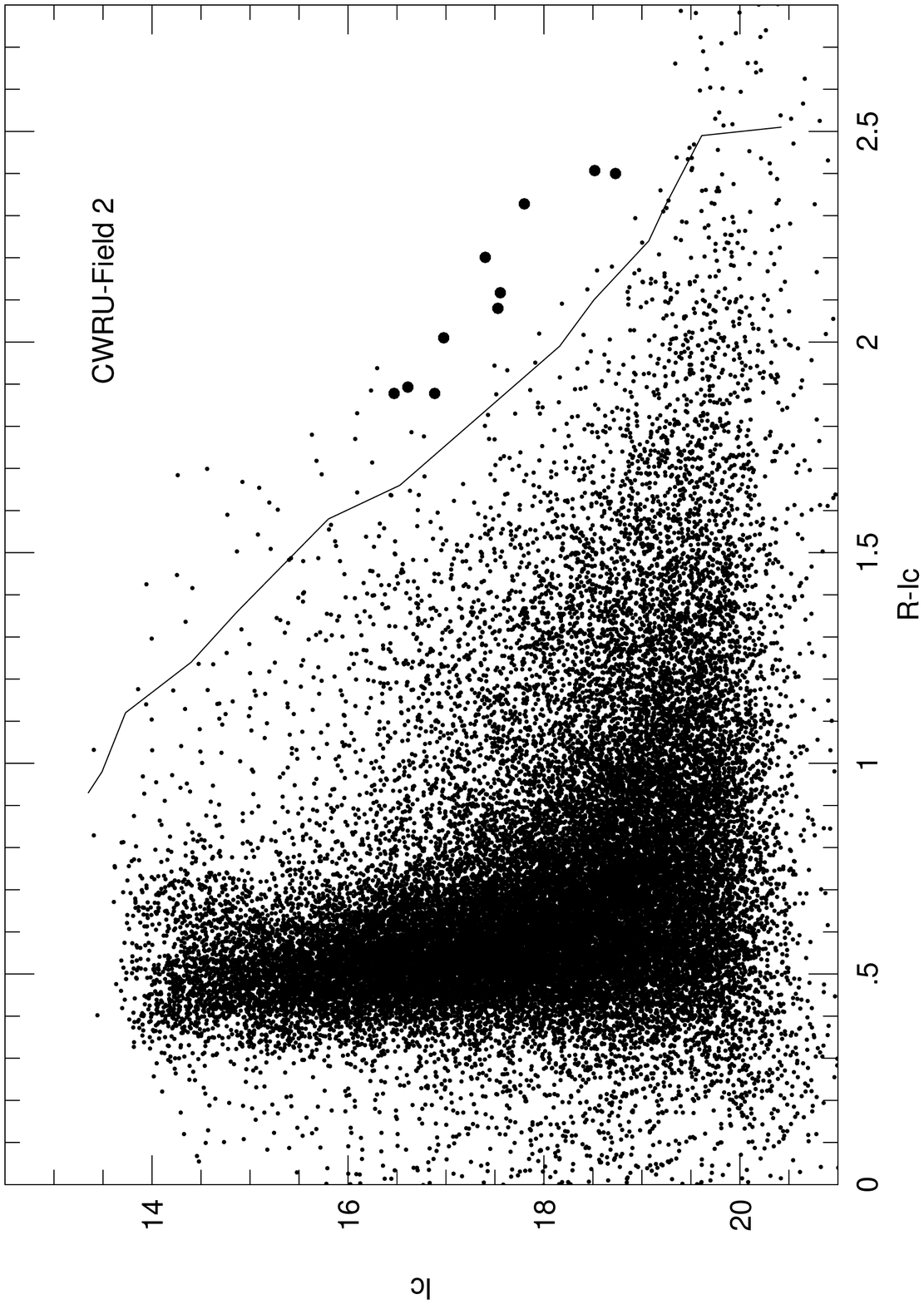}

\end{figure*}

\begin{figure*}
\vspace{18cm}

\includegraphics{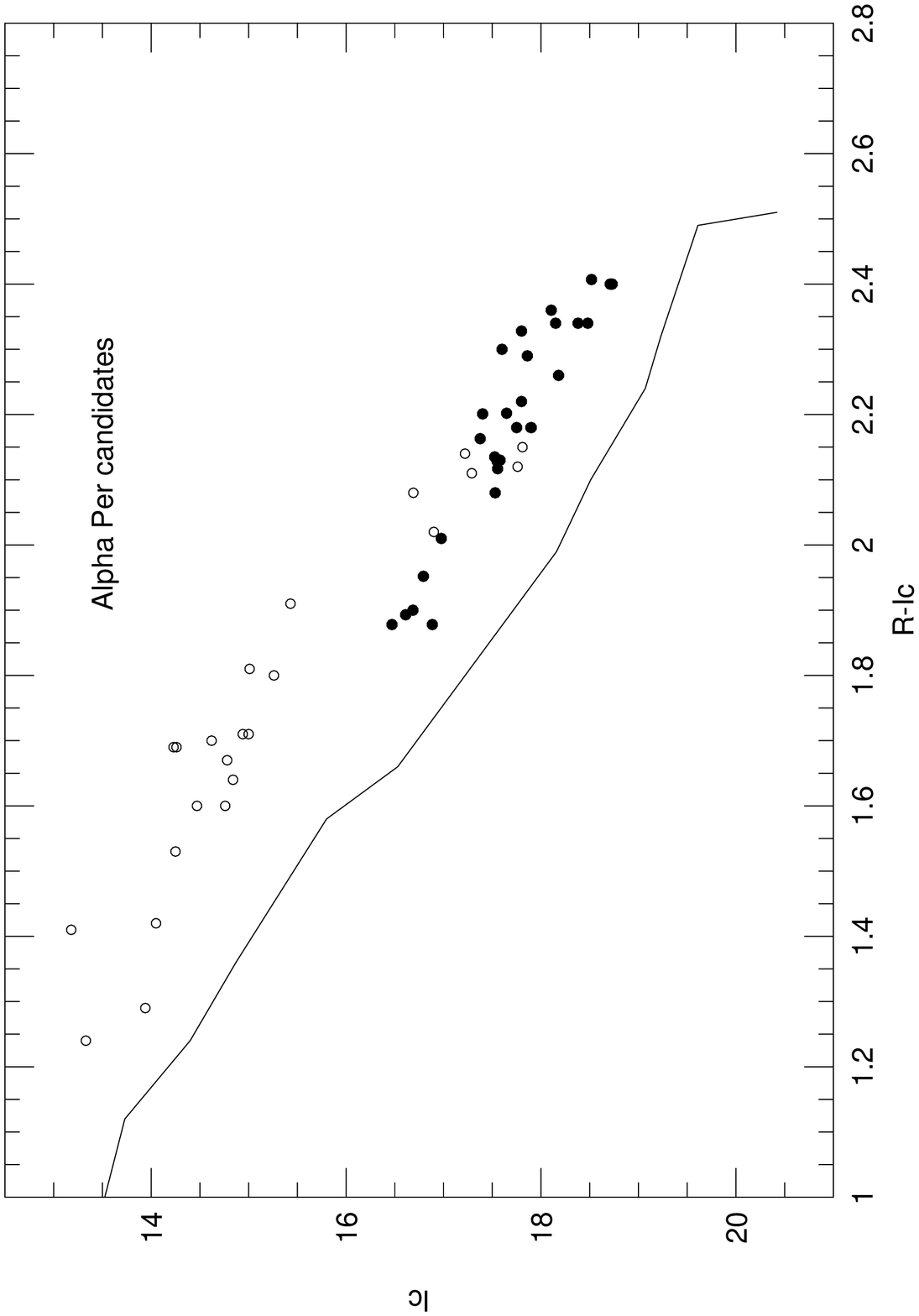}

\end{figure*}

\begin{figure*}
\vspace{18cm}

\includegraphics{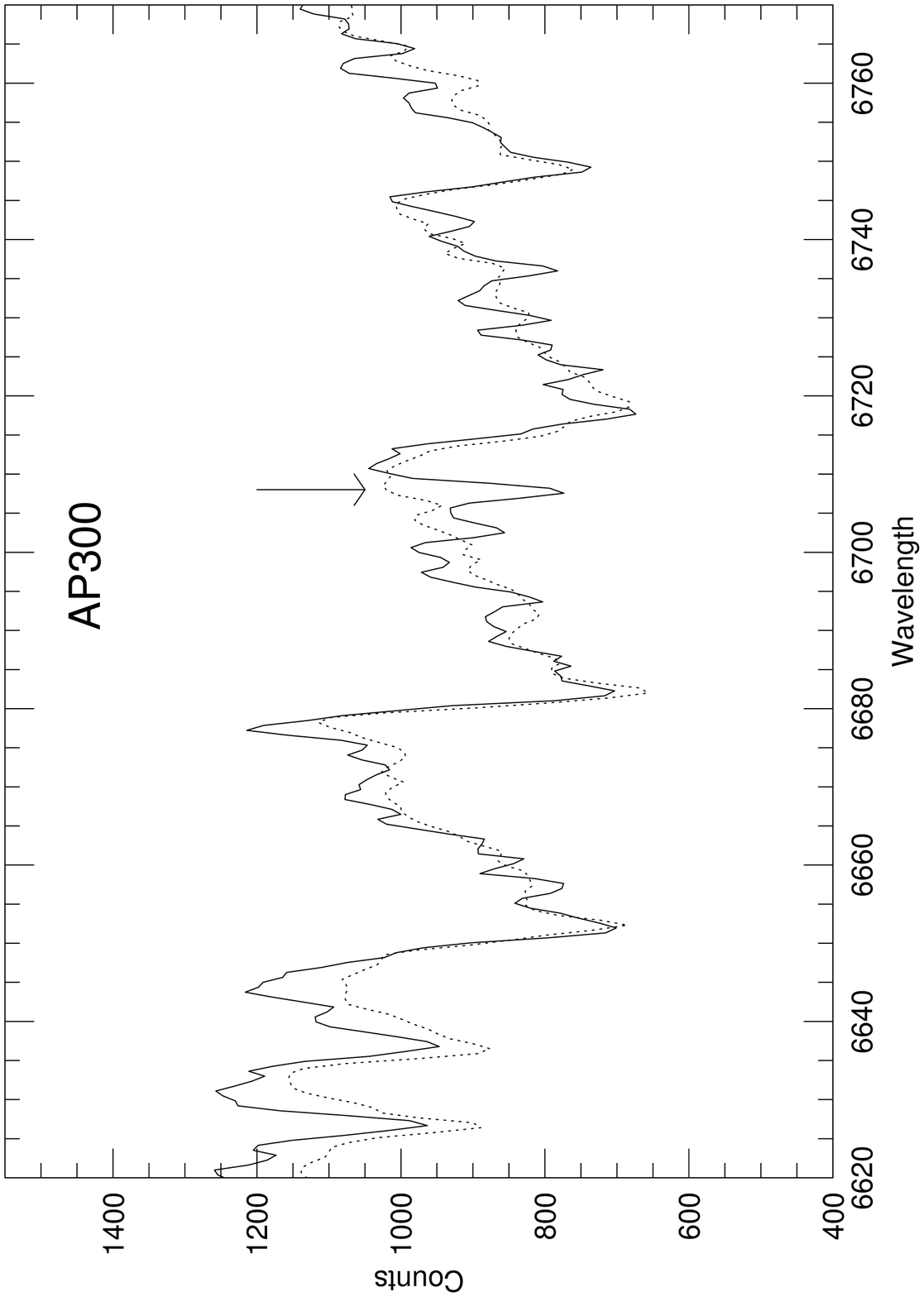}

\end{figure*}

\begin{figure*}
\vspace{18cm}

\includegraphics{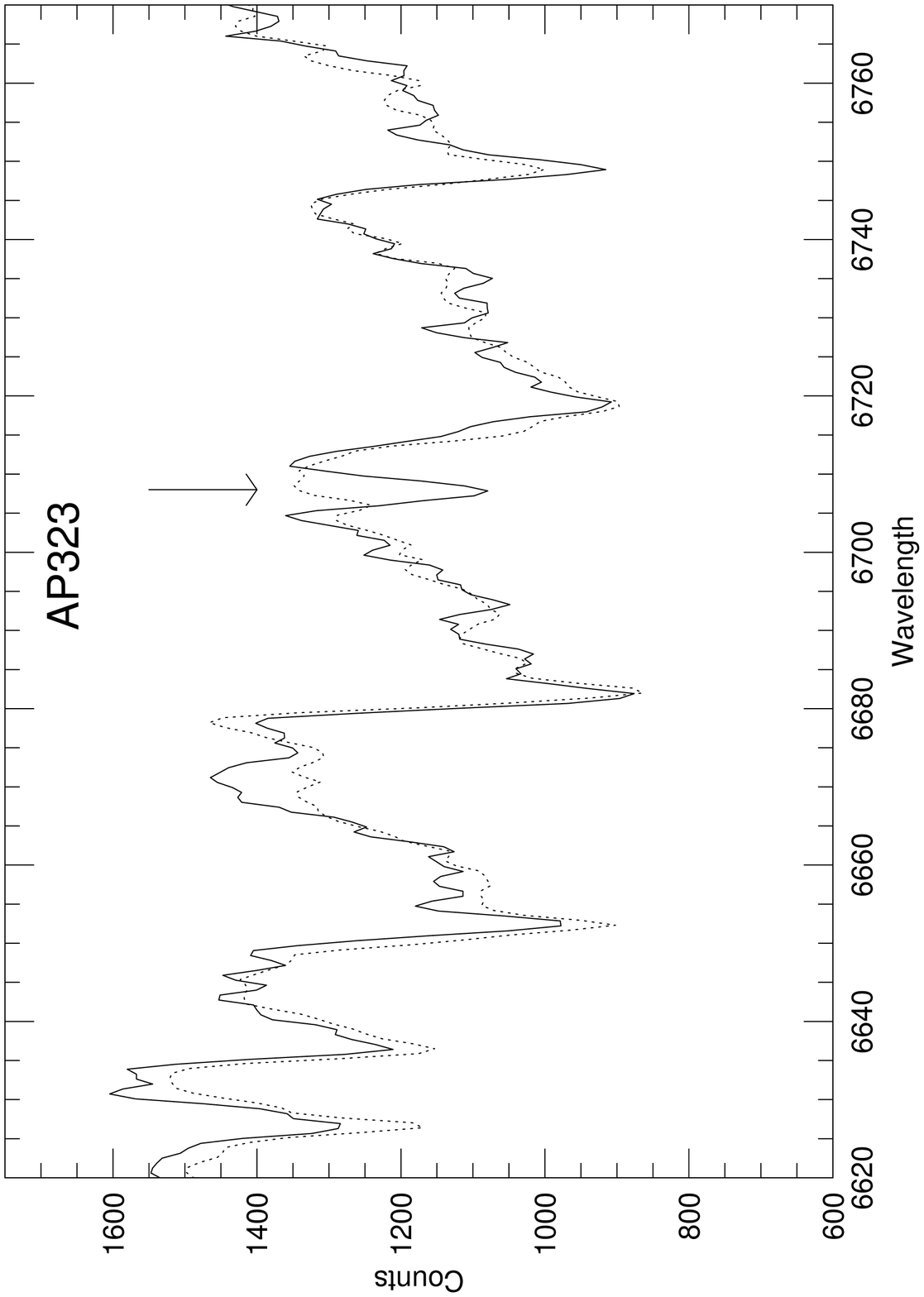}

\end{figure*}

\begin{figure*}
\vspace{18cm}

\includegraphics{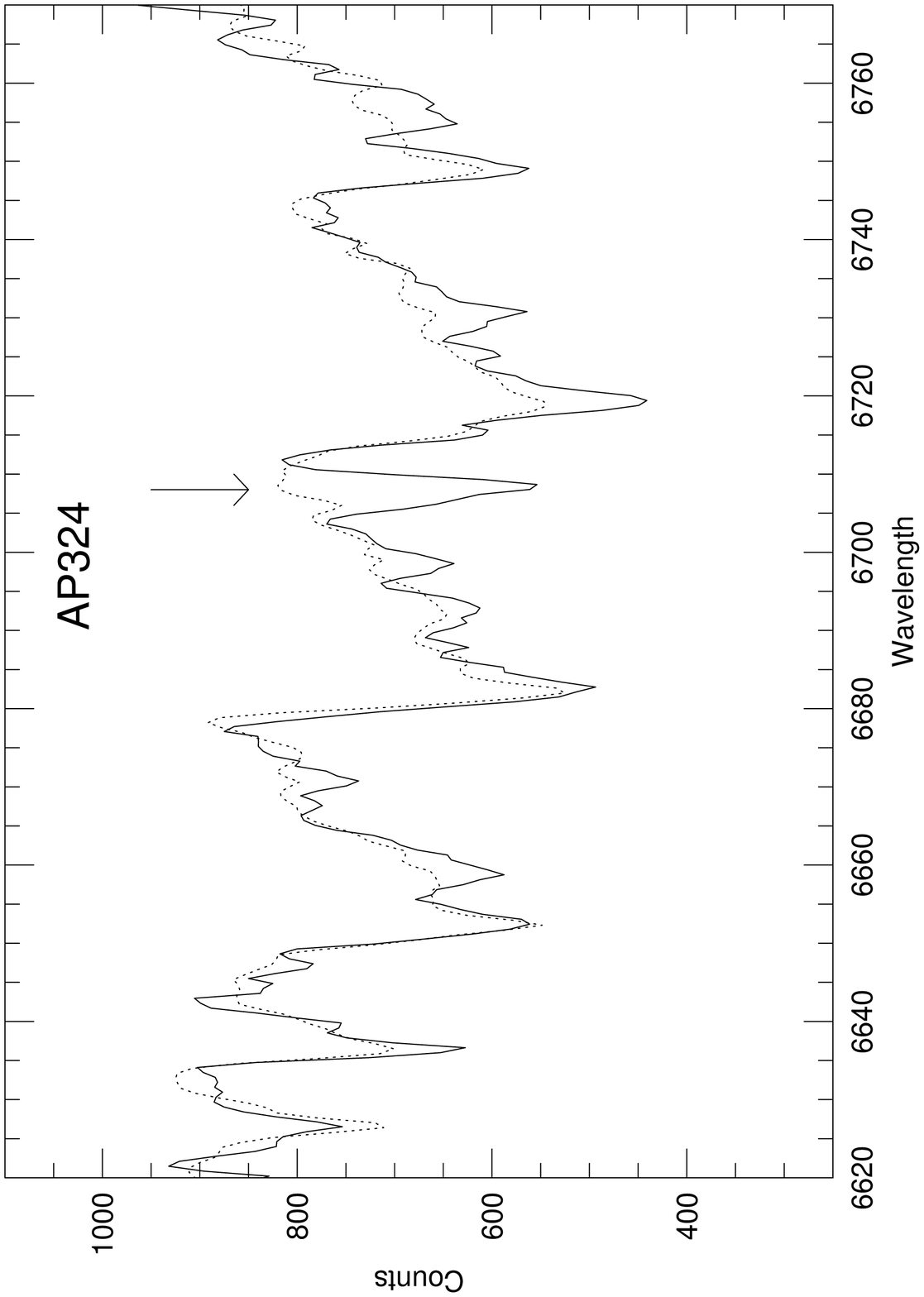}

\end{figure*}

\begin{figure*}
\vspace{18cm}

\includegraphics{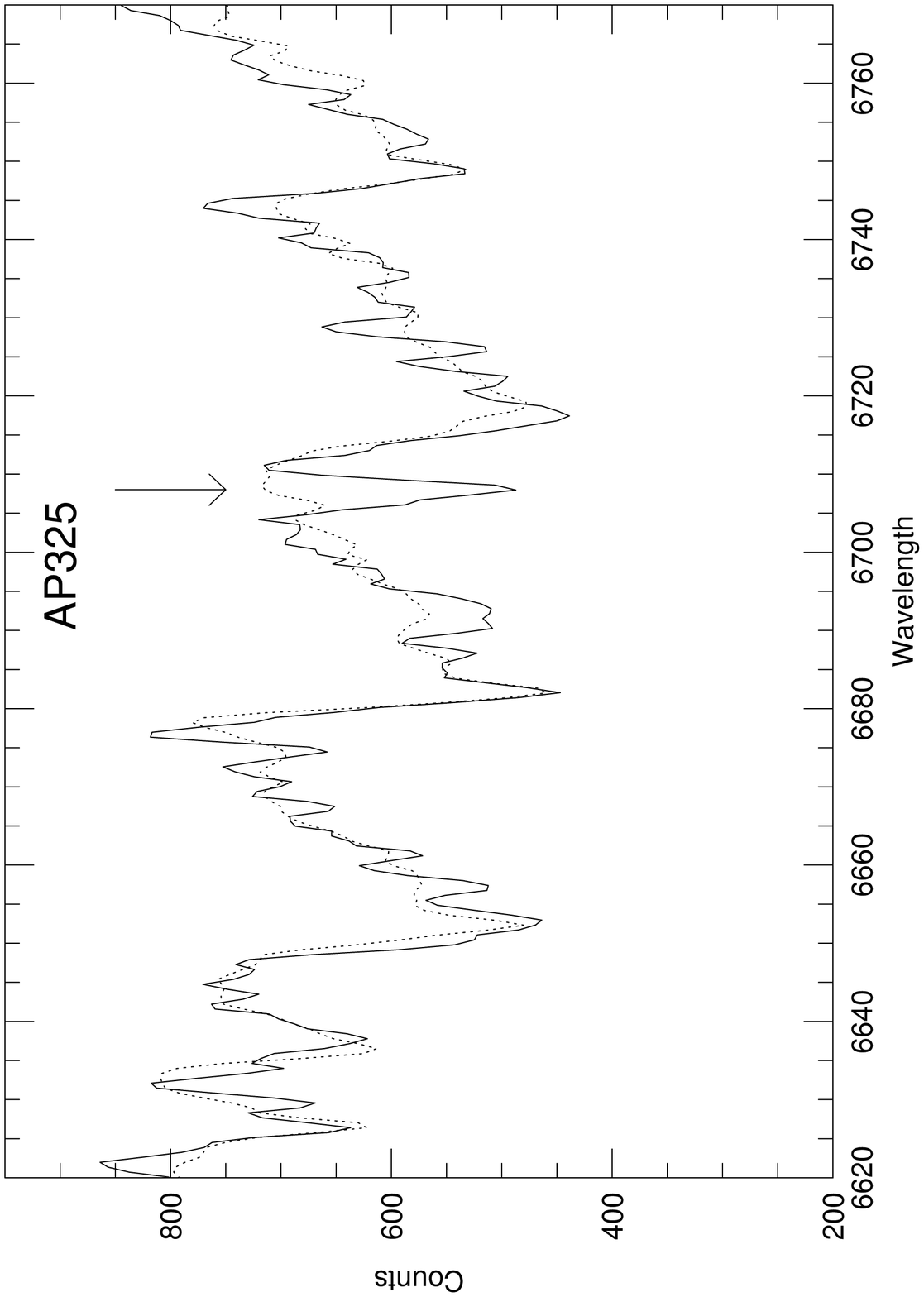}

\end{figure*}

\begin{figure*}
\vspace{18cm}

\includegraphics{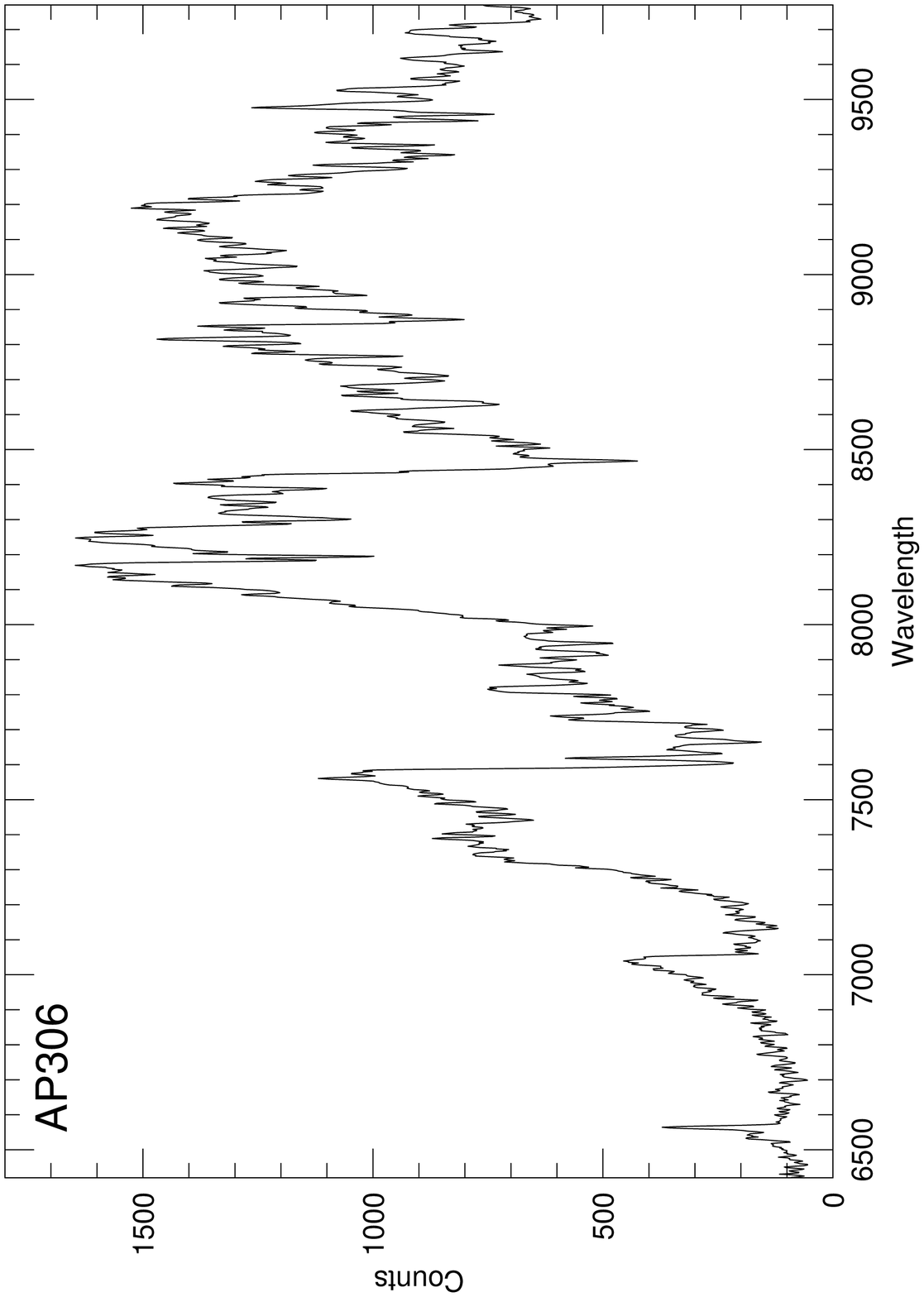}

\end{figure*}

\begin{figure*}
\vspace{18cm}

\includegraphics{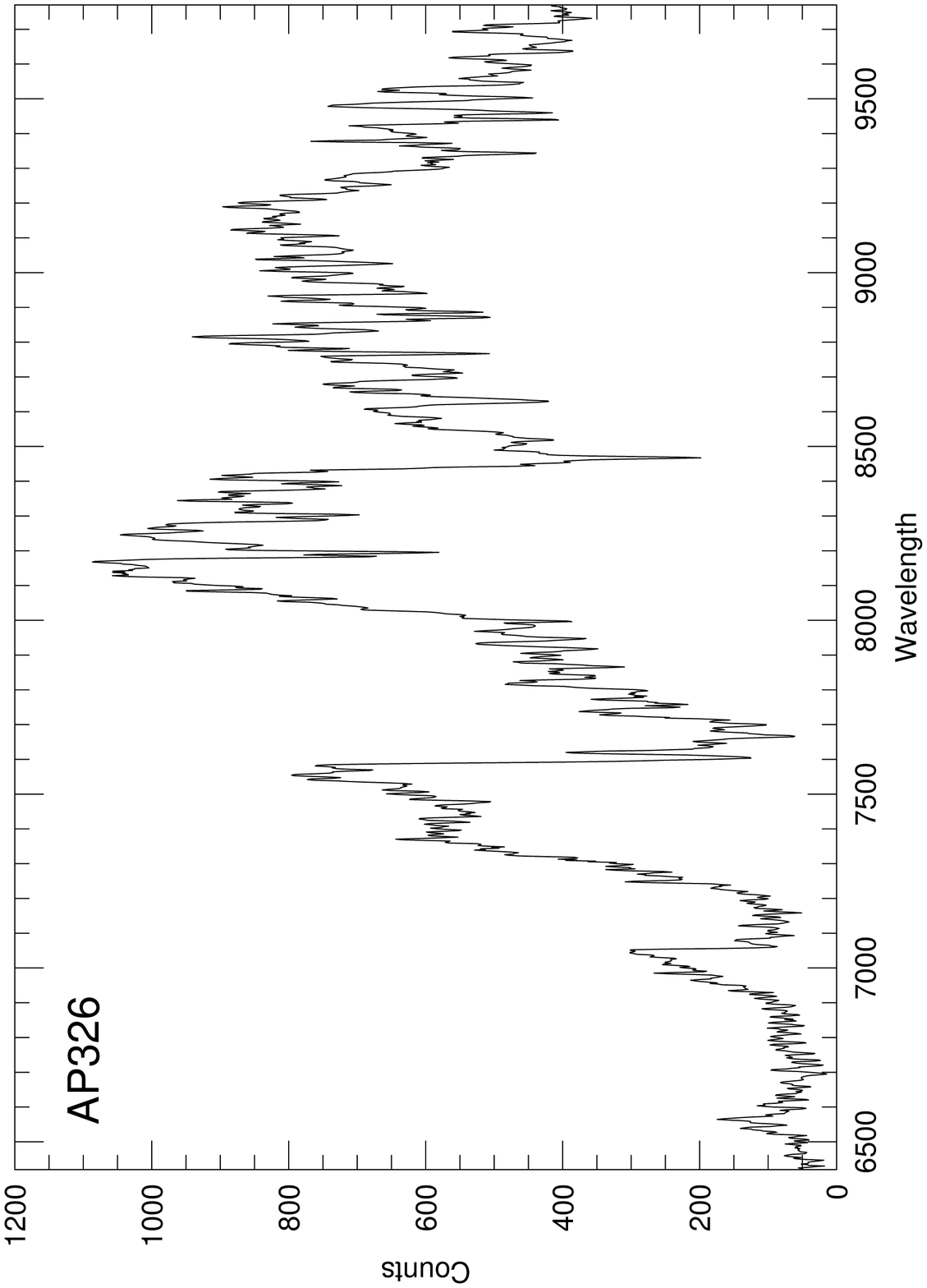}

\end{figure*}

\begin{figure*}
\vspace{18cm}

\includegraphics{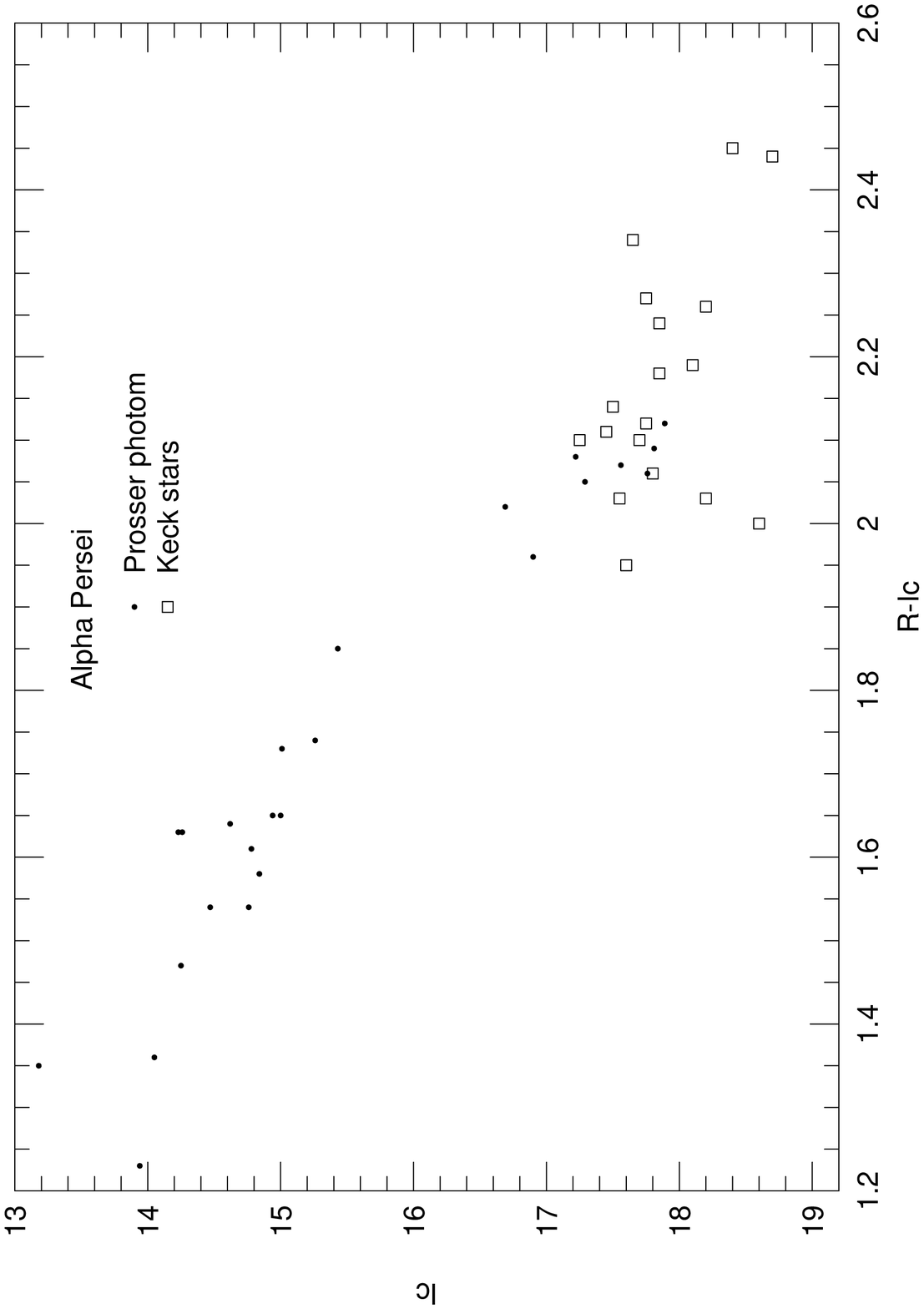}

\end{figure*}

\begin{figure*}
\vspace{18cm}

\includegraphics{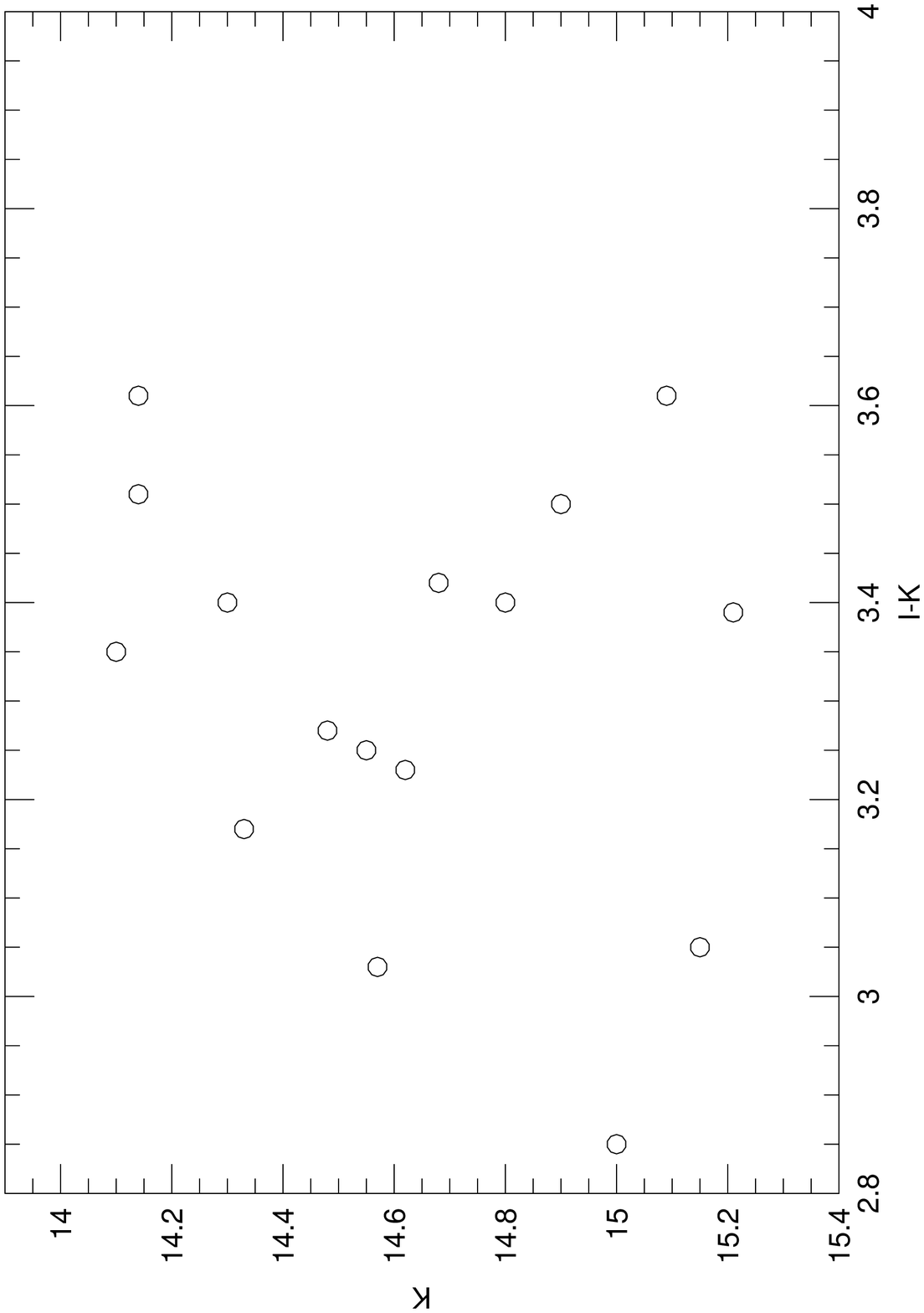}

\end{figure*}

\begin{figure*}
\vspace{18cm}

\includegraphics{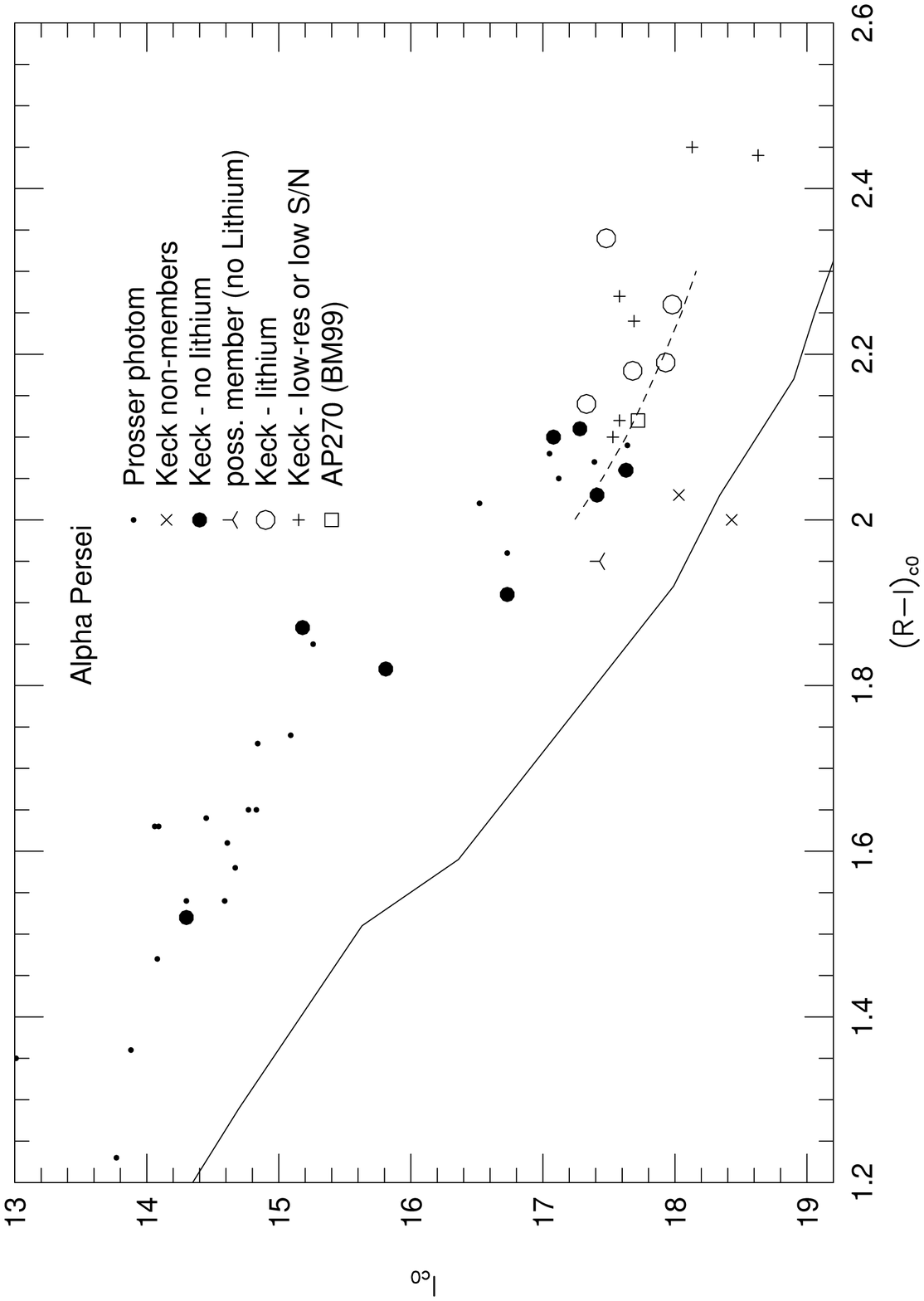}

\end{figure*}

\begin{figure*}
\vspace{18cm}

\includegraphics{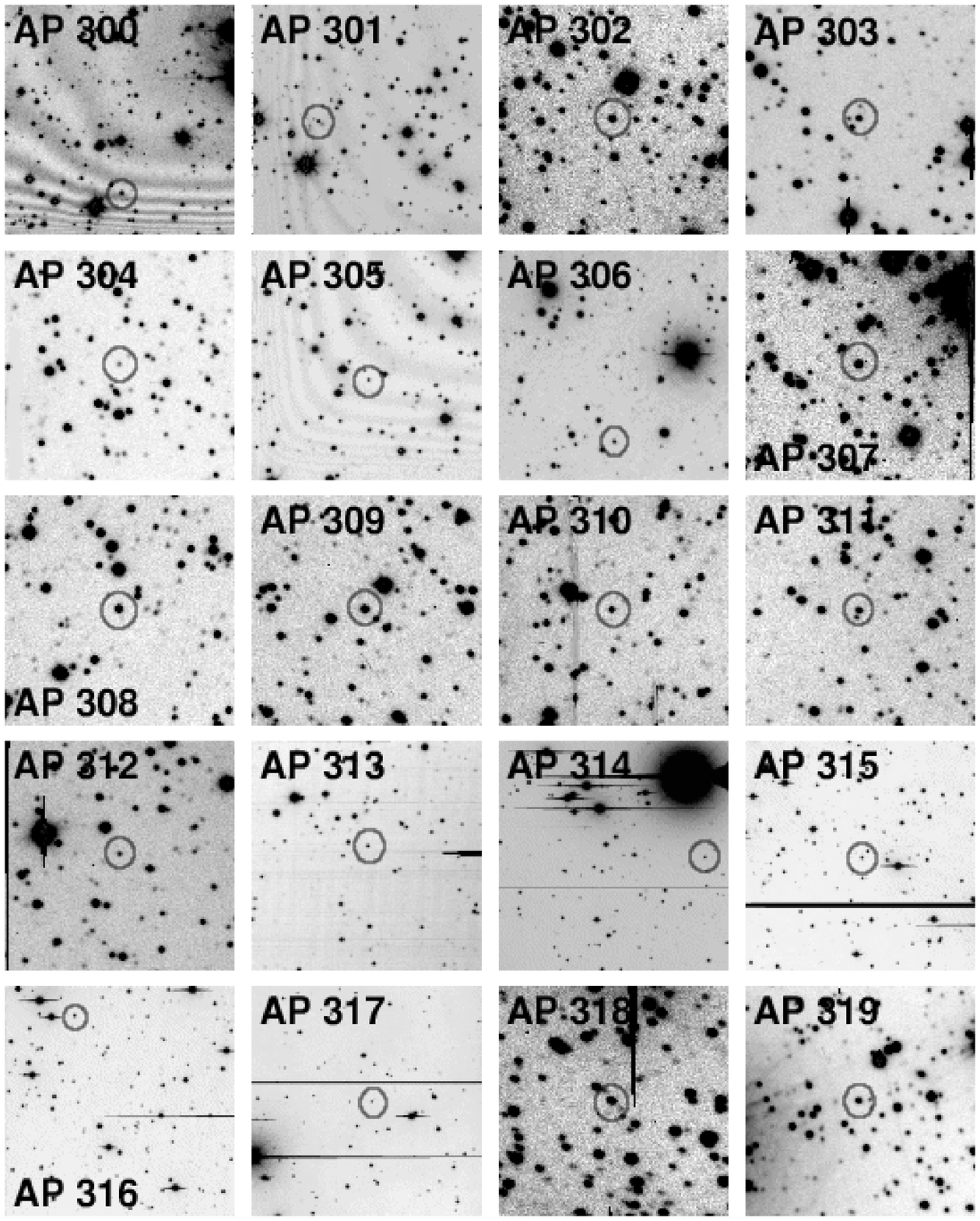}

\end{figure*}

\begin{figure*}
\vspace{18cm}

\includegraphics{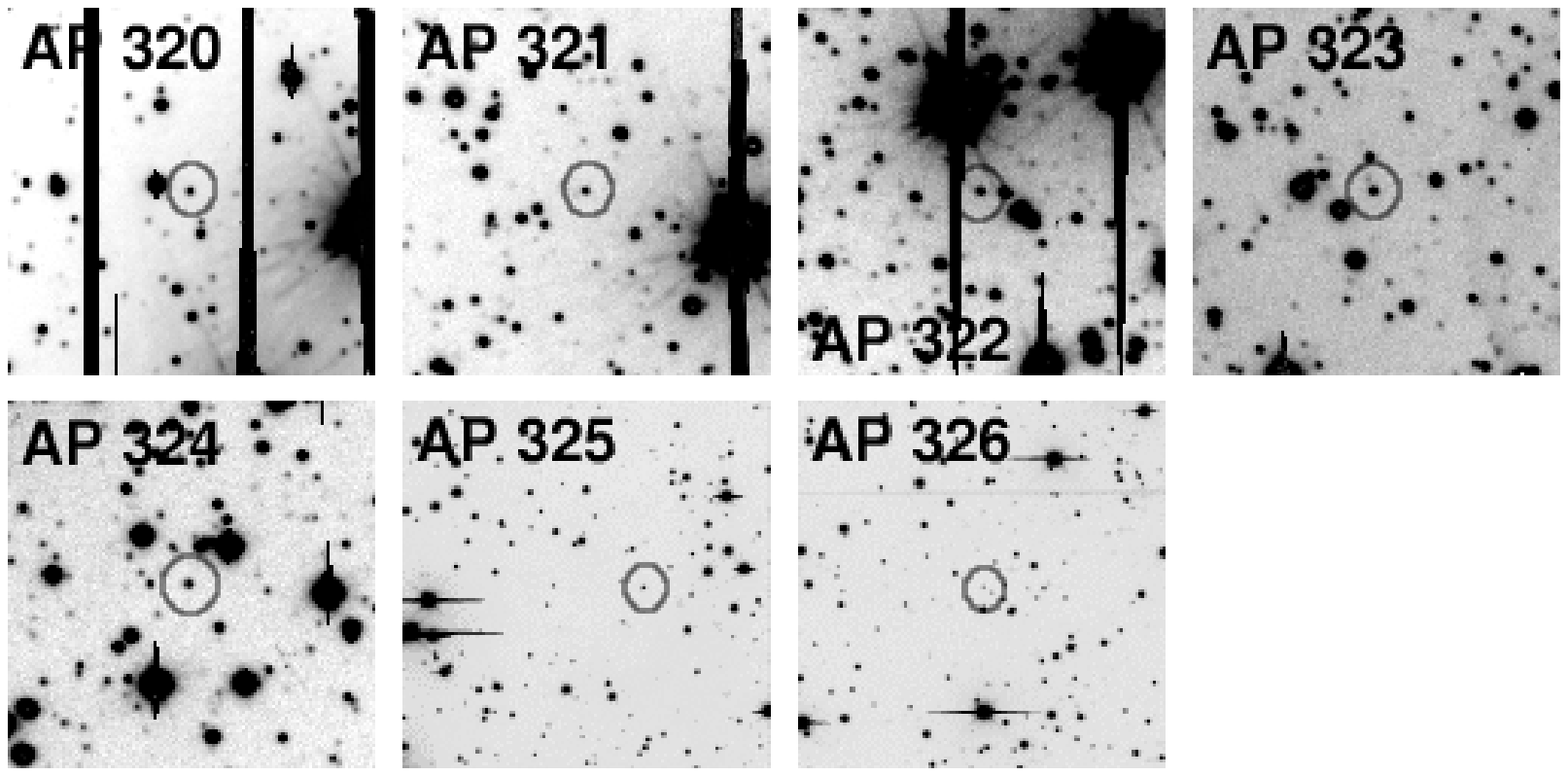}

\end{figure*}

\end{document}